# Quantum antiferromagnet bluebellite comprising a maple-leaf lattice made of spin-1/2 $Cu^{2+}$ ions

Yuya Haraguchi[1, *], Akira Matsuo[2], Koichi Kindo[2], and Zenji Hiroi[2]
[1]Department of Applied Physics and Chemical Engineering, Tokyo University of Agriculture and Technology, Koganei, Tokyo 184-8588, Japan
[2]The Institute for Solid State Physics, The University of Tokyo, Kashiwa, Chiba 277-8581, Japan
*chiyuya3@go.tuat.ac.jp

**Abstract**

Spin-1/2 maple leaf lattice antiferromagnets are expected to show interesting phenomena originating from frustration effects and quantum fluctuations. We report the hydrothermal synthesis of a powder sample of bluebellite $Cu_6IO_3(OH)_{10}Cl$ as a first potential candidate. Magnetization and heat capacity measurements reveal moderate frustration with a Curie–Weiss temperature of –35 K, and a magnetic transition at $T_N$ = 17 K. Surprisingly, the magnetic susceptibility and heat capacity above $T_N$ are well reproduced by the Bonner–Fisher model, which suggests that a one-dimensional spin correlation with a magnetic interaction of 25 K occurs in the apparently two-dimensional lattice. This emergent one-dimensionality cannot be explained by orbital ordering or dimensional reduction due to geometrical frustration. We believe that there is an unknown mechanism to cause one-dimensionality in the spin-1/2 maple leaf lattice antiferromagnet.

## I. Introduction

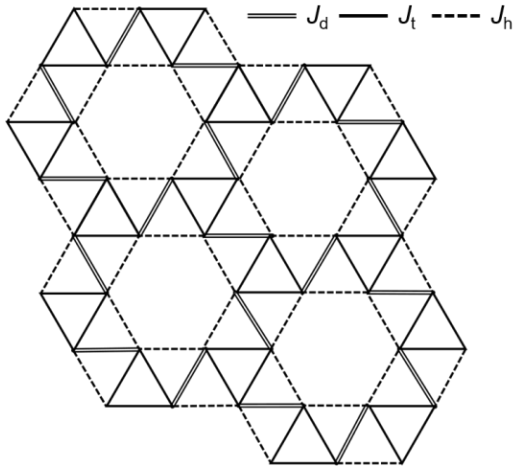

**Fig. 1** Maple-leaf lattice with the three types of magnetic interaction $J_d$, $J_t$, and $J_h$ for the dimers, triangles, and hexagons, respectively.

Over several decades, quantum spin frustrated magnets have attracted much attention from theorists and experimentalists searching for exotic states of matter such as the resonating valence bond state and spin liquids [1–4]. Synthetic copper minerals are candidates for such models. For example, compounds containing $Cu^{2+}$ ions with spin-1/2 in the kagomé net, such as herbertsmithite $Cu_3Zn(OH)_3Cl$ [5,6], volborthite $Cu_3V_2O_7(OH)_2 \cdot 3H_2O$ [7,8], vesignieite $BaCu_3V_2O_8(OH)_2$, [9,10], and engelhauptite $KCu_3V_2O_7(OH)_2Cl$ [11] have been extensively studied. These kagomé copper minerals have crystal structures based on layered hydroxides $M(OH)_2$. The magnetic ions $M$ are arranged in a triangular lattice, and 1/4 of them are periodically depleted or replaced by nonmagnetic ions to form a kagomé lattice.

There are other ways to generate characteristic frustrated lattices from a triangular lattice via periodic depletions. For example, a honeycomb lattice can be attained by a 1/3 depletion, and a maple-leaf lattice (MLL) can be attained by a 1/7 depletion [12–14]. As depicted in Fig. 1, all the lattice points in an MLL are equivalent, and each has five neighbors, similar to a natural maple leaf. When one type of magnetic ion is placed on the lattice points of the MLL, there are three kinds of magnetic interactions between them: $J_d$, $J_t$, and $J_h$ for the dimer, triangle, and hexagon, respectively. The ground state of the MLL antiferromagnet was predicted to have classical six-sublattice long-range order (LRO), even for quantum models [12,13]. However, more detailed theoretical investigations have shown that the $S$ = 1/2 MLL antiferromagnet exhibits a quantum dimer singlet state without LRO for $J_d/J \geq 1.45$ ($J = J_h = J_t$) [14]. Furthermore, a 1/3 plateau state may appear in the magnetization process for $J_d/J \geq 1.07$, and 1/3 and 2/3 plateaus may appear for $J_d/J \geq 3$ [14]. Thus, the MLL antiferromagnet is a potentially fascinating platform for the study of quantum frustrated magnetism, similar to kagomé antiferromagnets.

Unlike kagomé antiferromagnets, there are limited number of candidates for the MLL antiferromagnet, so that experimental studies have been limited. The magnetic properties associated with the chirality of the $J_t$ triangle have been reported for classical spin systems containing $Mn^{4+}$ ($S$ = 3/2) [15–17]. For $S$ = 1/2, several natural copper minerals may be candidates. Magnetization measurements have been performed on spangolite $Cu_6Al(SO_4)(OH)_{12}Cl_3 \cdot 3H_2O$ [18], but the data are obscured by the presence of impurity spins in a sample from natural mineral. In order to study the intrinsic magnetism of MLL antiferromagnets, a purified synthetic mineral is necessary. In this study, we successfully synthesized polycrystalline samples of bluebellite $Cu_6IO_3(OH)_{10}Cl$ as a candidate for the spin-1/2 MLL antiferromagnet by a hydrothermal method.

Bluebellite was discovered in 1983 in the Blue Bell Mine in the Mojave Desert, California. The crystal structure of bluebellite is based on a layered hydroxide $M(OH)_2$ structure containing $Cu^{2+}$ and nonmagnetic $I^{5+}$ ions in a 6:1 ratio in the two-dimensional (2D) triangular lattice plane forming an $S$ = 1/2 MLL (Fig. 2) [19]. Previous structural analysis using a natural mineral sample showed that the $Cu^{2+}$ and $I^{5+}$ ions are perfectly ordered without intersite mixing.

This work reports the crystal and magnetic properties of bluebellite. Magnetic susceptibility shows that it is an antiferromagnet with $S = 1/2$ and a Weiss temperature of $-35.6$ K. The magnetic susceptibility and heat capacity reveal magnetic ordering at $T_N = 17$ K. Both the temperature dependences in magnetic susceptibility and heat capacity above $T_N$ are well reproduced by the $S = 1/2$ Heisenberg chain model with $J \sim 25$ K in spite of the apparent two-dimensionality in magnetic couplings in the MLL. Moreover, a large $T$-linear term is observed in heat capacity above $T_N$, which is consistent with a one-dimensional (1D) spinon excitation with $J = 25.9$ K. Furthermore, the observed nonlinear increase in the magnetization process is similar to that observed in quantum spin chains. These observations strongly suggest that 1D spin correlations emerge in the 2D MLL of bluebellite owing to frustration and quantum effects.

## II. Experiments

A polycrystalline bluebellite sample was prepared using the hydrothermal method. 0.3 g of $Cu(OH)_2$, 0.3 g of $CuCl_2 \cdot 2H_2O$, and 0.1 g of $KIO_4$ were added to a polytetrafluoroethylene beaker of 30 ml volume. The beaker was placed in a stainless-steel vessel filled with 15 ml of $H_2O$ and heated at 150 °C for 8 h. A dark cyan-colored powder was obtained after rinsing with distilled water several times and drying at 110 °C. The obtained sample was characterized by powder X-ray diffraction (XRD) using a Cu-$K\alpha$ radiation. The cell parameters and crystal structure were refined by the Rietveld method using the RIETAN-FP version 2.16 software [20]. The temperature dependence of magnetization was measured under magnetic fields up to 7 T in a magnetic property measurement system (MPMS; Quantum Design). The temperature dependence of heat capacity was measured by a conventional relaxation method in a physical property measurement system (PPMS; Quantum Design). Magnetization curves up to approximately 60 T were measured by the induction method in a multilayer pulsed magnet at the International Mega Gauss Science Laboratory in Institute for Solid State Physics.

## III. Results

### A. Crystal structure

Figure 2(a) shows a powder XRD pattern for the sample. All of the peaks are indexed to a hexagonal unit cell with $a = 8.3066(2)$ Å and $c = 13.2211(4)$ Å based on the $R3$ space group reported for a natural mineral [19]. The Rietveld analysis converged well with the crystal structure shown in Fig. 2(b) and the structural parameters listed in Table 1; contributions from H atoms were ignored in the structural refinement. As reported previously, site-mixing between Cu and I ions was not detected within the experimental error [19].

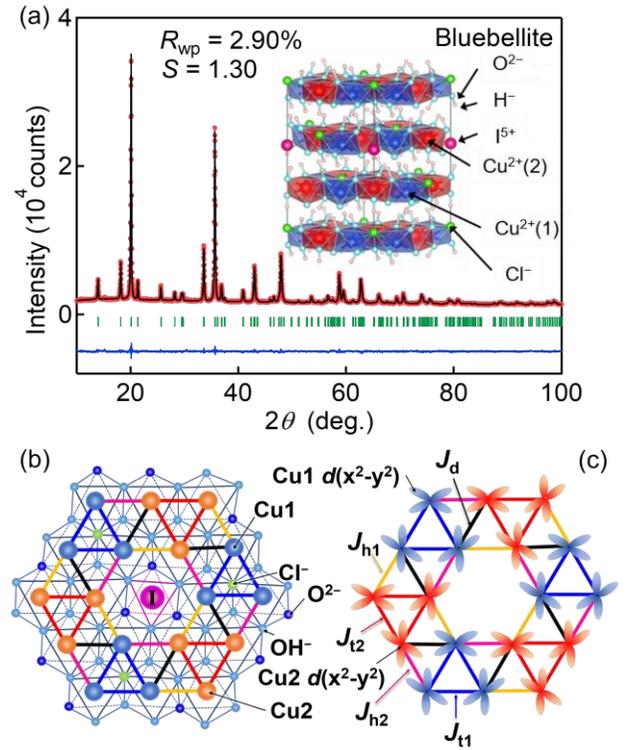

**Fig. 2** (Color online) (a) Powder X-ray diffraction pattern for synthetic bluebellite. The observed intensity, calculated intensity based on Rietveld analysis, and their difference are represented by red dots, black lines, and blue lines, respectively. The inset shows the crystal structure of bluebellite viewed along the layers visualized using the VESTA program [21]. (b) Arrangement of atoms in a layer. (c) Arrangement of spin-carrying Cu-$3d(x^2-y^2)$ orbitals. Blue and red lobes represent the $d(x^2-y^2)$ orbitals on the Cu1 and Cu2 sites, respectively. Five types of nearest-neighbor interaction are shown: $J_{t1}$ (blue), $J_{t2}$ (red), $J_d$ (black), $J_{h1}$ (green), and $J_{h2}$ (pink).

As mentioned in the Introduction, all of the magnetic ions in the MLL are identical to their five nearest neighbors and there are three different interaction paths. However, the space group of bluebellite is $R3$ without inversion symmetry, different from the ideal MLL ($R\bar{3}$), so that there are two Cu sites, each of which form a set of equilateral triangles with a different bond distance. In Fig. 2, these are depicted by blue and red triangles for sites Cu1 and Cu2, respectively. As a result, there are five types of bonds with different magnetic interactions: $J_{t1}$ (blue) and $J_{t2}$ (red) represent the Cu1–Cu1 and Cu2–Cu2 interactions on the corresponding triangles, respectively, and $J_d$ (black), $J_{h1}$ (green), and $J_{h2}$ (pink) represent the Cu1–Cu2 interactions, as shown in Fig. 2(c). Note that the $J_{h1}$ and $J_{h2}$ bonds appear alternately along the distorted hexagon.

**Table 1** Crystallographic parameters for synthetic bluebellite (space group $R3$) determined by Rietveld analysis of powder XRD data obtained at room temperature. The atomic coordinates and isotropic thermal displacement parameter $B$ are given for each atom. The refinement converged with small reliable parameters of $R_{wp} = 2.901\%$ and $S = 1.3004$. The obtained lattice parameters are $a = 8.3056(2)$ Å and $c = 13.2194(3)$ Å.

| Atom | Site | $x$ | $y$ | $z$ | $B$ (Å$^2$) |
|---|---|---|---|---|---|
| Cu1 | 9b | 0.4578(9) | 0.3867(9) | 0.2901(10) | 1.18(5) |
| Cu2 | 9b | 0.0261(6) | 0.2404(9) | 0.2747(11) | 0.84(5) |
| I | 3a | 0 | 0 | 0.6050(11) | 0.824(4) |
| O1 | 9b | 0.4178(29) | 0.1834(27) | 0.2212(11) | 0.22(9) |
| O2 | 3a | 0 | 0 | 0.2666(11) | 0.60(7) |
| O3 | 9b | 0.2571(35) | 0.4455(26) | 0.1999(11) | 0.45(10) |
| O4 | 9b | 0.2043(21) | 0.1939(22) | 0.3573(12) | 0.66(11) |
| O5 | 9b | 0.4521(31) | 0.1683(37) | −0.0044(10) | 0.82(12) |
| Cl | 3a | 0 | 0 | 0.0998(11) | 1.50(14) |

**Table 2** Cu-$X$ ($X$ = OH, O, Cl) bond lengths in bluebellite obtained from the powder XRD data.

| Bond | Length (Å) | Short/long |
|---|---|---|
| Cu1–O1H | 2.07(3) | Short |
| Cu1–O1H | 1.88(3) | Short |
| Cu1–O3 | 2.42(4) | Long |
| Cu1–O4H | 2.16(2) | Short |
| Cu1–O5H | 1.89(4) | Short |
| Cu1–Cl | 2.68(2) | Long |
| Cu2–O1H | 1.86(2) | Short |
| Cu2–O2H | 2.02(2) | Short |
| Cu2–O3 | 2.10(2) | Long |
| Cu2–O4H | 2.03(3) | Short |
| Cu2–O4H | 2.17(3) | Long |
| Cu2–O5H | 2.09(3) | Short |

Each Cu$^{2+}$ ion is surrounded by six anions: one Cl, one O, and four OH ions for Cu1 and one O and five OH ions for Cu2. For Cu1, the bonds are long for the Cl and O ions at the trans positions, and short for the four lateral OH ions. This produces 4-short–2-long type coordination with a large Jahn–Teller distortion (Table 2). In contrast, for Cu2, the O3 and O4H at the trans positions are located slightly further from Cu2, with the others ions closer. Therefore, it is likely that, for both Cu1 and Cu2, the $d(x^2-y^2)$ orbitals that extend toward the closest OH ions have the highest energy and thus carry spin-1/2. The arrangement of the $d(x^2-y^2)$ orbitals in the MLL layer derived from the crystal structure is shown in Fig. 2(c). Note that all superexchange interactions occur via OH$^-$ ions located between two lobes of nearby orbitals; the Cl ion above the blue Cu1 triangle and the O$^{2-}$ ion surrounding the I$^-$ ion do not participate.

B. Physical properties

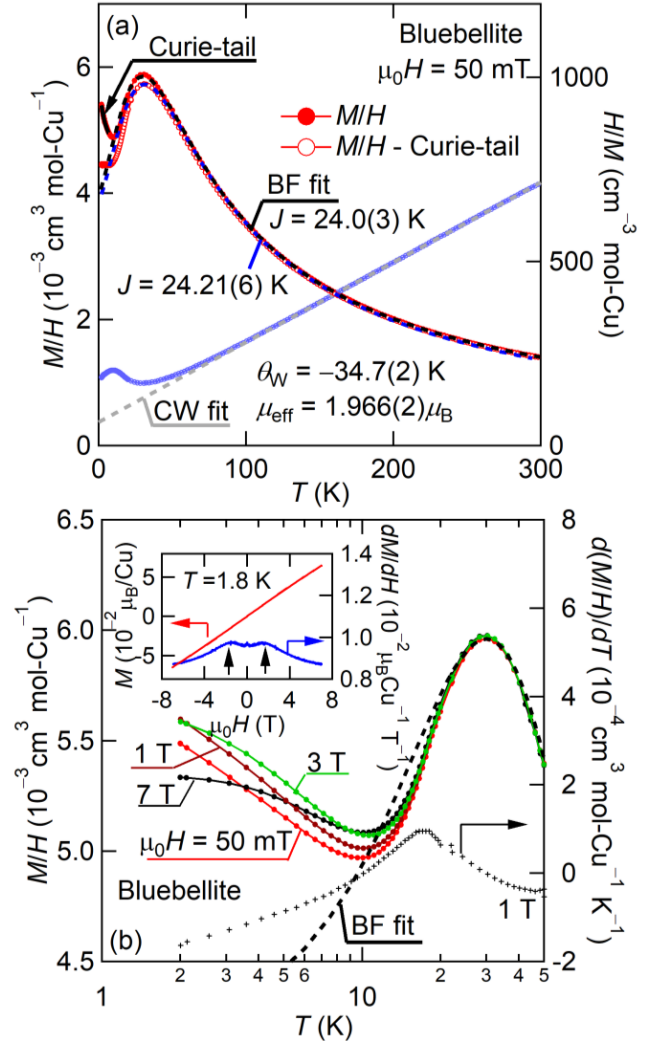

**Fig. 3** (Color Online) (a) Temperature dependence of the magnetic susceptibility $M/H$ and its inverse for a polycrystalline sample of bluebellite measured under an applied magnetic field of 50 mT. The $M/H$ data after the subtraction of the low-temperature Curie-tail contribution is also shown. The dashed lines on the $M/H$ and $H/M$ data represent fits to the Bonner–Fisher (BF) [23] and Curie–Weiss (CW) models, respectively. (b) Temperature dependences of $M/H$ measured at magnetic fields of 0.05, 1, 3, and 7 T. Magnetic ordering is observed at $T_N = 17$ K, where the derivative of the $M/H$ curve at 1 T shows a peak and field dependence starts to appear. The inset shows the $M$–$H$ curve (red) and its derivative (blue) measured at 1.8 K. The black arrows indicate spin flop transitions at ±1.7 T.

The temperature dependences of magnetic susceptibility $M/H$ and its inverse $H/M$ are shown in Fig. 3(a). A Curie–Weiss (CW) fit of the $H/M$ curve in the 200–300 K range yields a Weiss temperature of $\theta_W = -34.7(2)$ K and an effective magnetic moment $\mu_{\text{eff}}$ of $1.966(2)\mu_B$. The negative $\theta_W$ value indicates that the interaction between Cu$^{2+}$ spins is predominantly antiferromagnetic, and the $\mu_{\text{eff}}$ value suggests an $S = 1/2$ spin system with Lande's $g$-factor of $g = 2.27$, which is enhanced by the spin-orbit interaction [22] and is comparable to those of other copper minerals [5–11]. The average magnetic interaction $J$ is given by ($J_{t1} + J_{t2} + J_d + J_{h1} + J_{h2}$)/5 in the mean-field approximation, and it is calculated to be 27.8 K, where $\theta_W = -zJS(S + 1)/3$ with $z = 5$ and $S = 1/2$.

In the low-temperature region of Fig. 3(a), the $H/M$ curve begins to deviate from the CW line below ~70 K, and the $M/H$ curve shows a broad peak at ~30 K, suggesting the development of antiferromagnetic short-range order (SRO). Interestingly, this broad peak is well reproduced by the Bonner–Fisher (BF) curve for a spin-1/2 Heisenberg antiferromagnetic chain with an intrachain interaction where $J = 24.0(3)$ K and $g = 2.27(3)$ [23]. This $J$ value is close to the average $J$ of 27.8 (3) K from the CW fit, and the $g$ values from the two fits are identical. This fact strongly suggests that the observed SRO has 1D characteristics. Note that there is an upturn below 10 K, which is ascribed to a contribution of impurity spins of ~1.5%. A change in the $M/H$ curve after the subtraction of this additional contribution was minimal, and a BF fit to the corrected data gave a slightly larger $J$ of 24.21(6) K.

Figure 3(b) shows the $M/H$ curves below 50 K measured under various magnetic fields. The derivative of the curve at 1 T shows a peak at an inflection point at 17 K. Moreover, a field dependence appears below this temperature. In contrast, the isothermal magnetization curve at 1.8 K in the inset of Fig. 3(b) shows small increases at $\mu_0 H = \pm 1.7$ T, as evidenced by the peaks in its derivative $dM/dH$, suggesting the occurrence of spin-flop transitions. Thus, the magnetic field dependence of $M/H$ below 17 K is due to a spin-flop transition. These observations indicate that magnetic LRO occurs at $T_N = 17$ K. In fact, as shown in Fig. 4, the heat capacity divided by temperature $C/T$ exhibits a clear λ-shaped peak at 16.9 K under 0 and 7 T fields. Therefore, in bluebellite, a second-order magnetic phase transition with bulk nature occurs at $T_N$.

A plot of $C/T$ against $T^2$ is shown in the inset of Fig. 4, where $C/T$ varies linearly with $T^2$ over a wide temperature range from above $T_N$ to approximately 40 K; a large intercept of $\gamma = 0.214(3)$ J mol-Cu$^{-1}$ K$^{-2}$ is obtained by fitting to the equation $C/T = \gamma + \beta T^2$. In general, a cubic heat capacity term comes from phonon contributions, whereas a $T$-linear term comes from densely populated electronic excitations. Since bluebellite is an insulator, this $T$-linear term must originate from spins rather than conduction electrons. It is known that such a $T$-linear heat capacity appears from a spinon Fermi surface. A Tomonaga–Luttinger liquid, which is the ground state of a 1D spin-1/2 Heisenberg chain [24,25], is no example. Another possible origin is the gapless excitation in the spin glass. However, this is unlikely because no glassy responses were observed in the magnetic susceptibility, and a sharp transition to an LRO was observed in the heat capacity. Considering that the broad peak in the $M/H$ was reproduced by a BF curve, it is reasonable to assume that the observed $T$-linear heat capacity above $T_N$ comes from spin correlations of the 1D character. In this scenario, the magnetic interaction $J$ in the chain is given by $\gamma = 2R/3J$, where $R$ is the gas constant [23, 26]. Thus, we obtain $J = 25.9(3)$ K, which is close to $J = 24.0(3)$ K from the BF fitting to $M/H$.

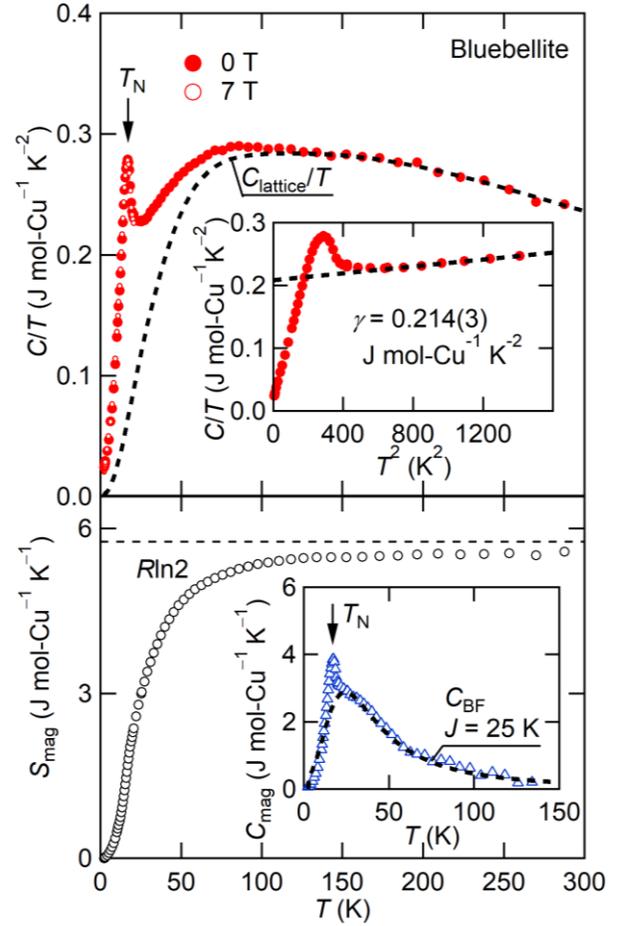

**Fig. 4** (Color Online) Temperature dependences of heat capacity divided by temperature $C/T$ and magnetic entropy $S_{mag}$ for bluebellite. Red circles (filled and open for $\mu_0 H = 0$ and 7 T, respectively) and dashed line represent the data and lattice contribution estimated by fitting the data above 100 K ($C_{lattice}/T$), respectively. The black arrow indicates the magnetic transition temperature $T_N$. The inset in the top panel shows a plot of $C/T$ against $T^2$ revealing the presence of a large $T$-linear contribution in $C$. Magnetic entropy $S_{mag}$ approaches $R\ln 2$ at high temperatures. The inset in the bottom panel shows the magnetic heat capacity $C_{mag}$ after the subtraction of $C_{lattice}$. The broken line is a fit to the BF model, which yields $J = 25$ K.

The magnetic heat capacity $C_{mag}$ has been estimated by calculating the lattice heat capacity $C_{lattice}$ and subtracting it from the total $C$. It is known that $C_{lattice}$ consists of contributions from three acoustic phonon branches and ($3n - 3$) optical phonon branches, where $n$ is the number of atoms per formula unit [27]; $n$ equals 21 for bluebellite. Provided that $C_{lattice}$ is the sum of Debye- and Einstein-type heat capacities, $C_D$ and $C_E$, respectively, the $C/T$ data above 150 K were fitted to the equation,

$$C_{lattice} = C_D + C_E$$
$$= 9R(T/\theta_D)^3 \int_0^{\theta_D/T} \frac{x^4 \exp(x)}{[\exp(x)-1]^2} dx$$
$$+ R \sum_{i=1}^{3} n_i \frac{(\theta_{Ei}/T)^2 \exp(\frac{\theta_{Ei}}{T}-1)}{\exp(\frac{\theta_{Ei}}{T}-1)}, \quad (2)$$

where $R$ is the gas constant, $\theta_D$ is the Debye temperature, $\theta_{Ei}$ are the Einstein temperatures. The best fit shown by the dashed line in Fig. 4 was obtained assuming three kinds of Einstein modes: $\theta_D = 161$ K, $\theta_{E1} = 257$ K, $\theta_{E2} = 655$ K, $\theta_{E3} = 971$ K, $n_1 = 18$, $n_2 = 28$, and $n_3 = 14$. The $C_{mag}/T$ was obtained by subtracting the $C_{lattice}/T$ from the experimental data, and the magnetic entropy $S_{mag}$ was calculated by integrating $C_{mag}/T$ with respect to $T$. The obtained $S_{mag}$ above 100 K approaches the total entropy of spin 1/2, $R\ln 2 = 5.76$ J mol$^{-1}$ K$^{-1}$, indicative of the validity of the subtraction of lattice contributions. The magnetic entropy reaches 1.61 J mol$^{-1}$ K$^{-1}$ at $T_N$, which is 28% of the total entropy, indicating that short-range magnetic correlations already developed above $T_N$, corresponding to the broad peak in the magnetic susceptibility. Moreover, as shown in the bottom inset of Fig. 4, the thus-obtained $C_{mag}$ data above $T_N$ is well reproduced by the BF model with $J = 25$ K [28], which is reasonably in good agreement with $J = 24.0(3)$ K from the BF fit to the magnetic susceptibility and $J = 25.9(3)$ K from the $T$-linear heat capacity.

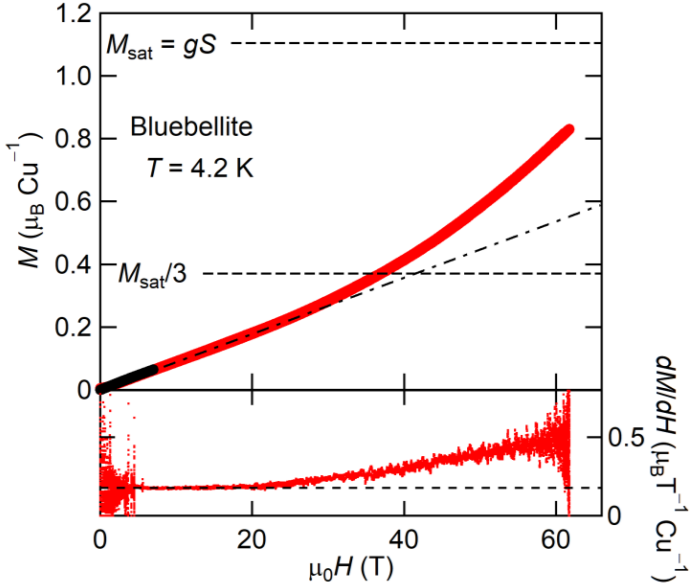

**Fig. 5** Magnetization curve and its derivative measured at 4.2 K under pulsed magnetic fields up to 61.7 T for bluebellite. The magnitudes are calibrated to the data measured under static fields up to 7 T (black circles). The saturated magnetization $M_{sat}$ is $gS$ with a $g$-value of 2.27, which was obtained from the CW fitting.

Magnetization measurements were performed under pulsed magnetic fields up to 61.7 T, as shown in Fig. 5. The magnetization $M$ increases almost linearly below 20 T, and then increases more rapidly to approximately 80% of the saturation magnetization $M_{sat} = gS$ at 61.7 T. No anomalies indicating a field-induced phase transition or a magnetization plateau are observed. Note that such a superlinear increase of magnetization is often observed in quantum spin chains [29].

## IV. Discussion

Bluebellite is an antiferromagnet with an $S = 1/2$ 2D MLL. However, unexpected 1D-like SRO are observed. The magnetic susceptibility and magnetic heat capacity show broad peaks at temperatures above $T_N$ that are well reproduced by the BF model for spin-1/2 Heisenberg antiferromagnetic chains. In addition, a large $T$-linear heat capacity is observed, which suggests the presence of gapless spin excitations, such as in the Tomonag–Luttinger liquid. The intra-chain interaction is consistently estimated from $M/H$ and $C/T$ data as approximately 25 K. These facts support that the 1D spin model is a good solution, although we cannot rule out other spin models.

We have to be careful in concluding the presence of 1D-like SRO in bluebellite because there are many quantum spin systems that show similar peaks in magnetic susceptibility [30]. We point out that the peak tends to be shallower for 2D than 1D systems [31]. An alternative way to explain the observed peak is to assume a spin-gapped state. In fact, we could fit the data assuming two gaps of 33.6 and 92.2 K. However, such a spin-gapped state is not consistent with the observed $T$-linear heat capacity without an exponential decay. It is emphasized that both the magnetic susceptibility and magnetic heat capacity data above $T_N$ are well reproduced by the BF model with nearly equal $J$ values of 25 K. Therefore, we think that there is an emergent 1D spin correlation in the MLL lattice of bluebellite.

Magnetic low dimensionality of spins is usually materialized by anisotropic chemical bonding in actual three-dimensional crystals. However, it can emerge beyond the topology of the crystal structure owing to anisotropic magnetic interactions caused by specific orbital arrangements and frustration effects. For example, in spinel oxides $ZnCr_2O_4$ and $ZnV_2O_4$, magnetic low-dimensionality occurs in three-dimensional pyrochlore lattices [32, 33]. In $ZnCr_2O_4$, the frustration effect causes the formation of zero-dimensional spin hexamers in the pyrochlore lattice in the excited state [32], while, in $ZnV_2O_4$, spin chains form owing to orbital ordering [33]. In contrast, dimensional reductions from 2D to 1D have been observed in anisotropic triangular lattice antiferromagnets, $Cs_2CuCl_4$ [34–36] and $A_3ReO_5Cl_2$ ($A$ = Ca, Sr, Ba) [37,38]. The one-dimensionalization of these compounds is caused by the geometric cancelation of magnetic interactions at the zigzag bonds between the chains. A further interesting dimensional reduction from 3D to 2D has been observed in a synthetic mineral pharmacosiderite $(H_3O)Fe_4(AsO_4)_3(OH)_4 \cdot 5.5H_2O$ consisting of $Fe^{3+}$ tetramers arranged in a primitive cubic lattice [39,40].

In the case of bluebellite, however, it seems difficult to explain the observed 1D magnetism based on these scenarios. In the orbital arrangement expected from the crystal structure of bluebellite shown in Fig. 2(c), there are no arrays of chains of orbitals. Moreover, even assuming that some of the $J$ couplings in the distorted MLL are negligible, it is impossible to generate 1D chains because of the presence of a three-fold rotation axis. Therefore, a novel mechanism is required to explain the 1D magnetism in bluebellite, which may incorporate both frustration and quantum effects. Recently, Makuta and Hotta suggested that the ground state of the $S = 1/2$ MLL has a 1D "striped" order, regardless of the ratio of $J$, based on finite-size exact diagonalization analysis of the ground states that constitute the low-energy states in the Heisenberg model [41]. It is explained that these one-dimensional magnetic structures are stabilized by the propagation of two spinons generated by the splitting of magnon in a one-dimensional direction along with the stripe. In this situation, they suggest that weakening $J_{t1}$ strengthens the one-dimensionality. In addition, they have also found that the calculated χ with smaller $J_{t1}$ can be fitted well with the 1D-XXZ spin chain model [41]. A similar emergence of 1D physics from 2D spins has been discussed for the distorted Shastry–Sutherland lattice [42]. Thus, it is plausible that one-dimensionality is hidden in the $S = 1/2$ MLL antiferromagnet, which will be clarified theoretically in the future.

It is necessary to accumulate more experimental evidence of 1D magnetism in bluebellite. Inelastic neutron scattering (INS) experiments are planned to estimate the exchange parameters needed to construct a reasonable model and reveal the nature of low-energy excitations. In addition, it would be interesting to examine whether 1D magnetism is also observed in other MLL copper minerals such as fuettererite $Pb_3Cu_6TeO_6(OH)_7Cl_5$ [43], mojaveite $Cu_6TeO_4(OH)_9Cl$ [19], spangolite $Cu_6Al(SO_4)(OH)_{12}Cl_3·3H_2O$ [18], and sabelliite $Cu_2ZnAsO_4(OH)_3$ [44]. Synthesis methods will be established for these minerals so that their magnetism may be elucidated.

## V. Summary

We synthesized the mineral bluebellite as a candidate for the $S = 1/2$ MLL antiferromagnet by a hydrothermal method and investigated its crystal structure and magnetic properties. Unexpected 1D-like SRO was observed in the magnetic susceptibility and heat capacity above $T_N$ which are reproduced by the BF model, the large $T$-linear heat capacity, and the nonlinear growth in the high-field magnetization in spite of the apparent 2D crystal structure. Our findings demonstrate that MLL copper minerals provide a promising platform in the search for novel quantum magnetism.

## Acknowledgments

We are grateful to R. Makuta and C. Hotta for their helpful discussions. This work was partially supported by the Core-to-Core Program for Advanced Research Networks provided by the Japan Society for the Promotion of Science (JSPS).

## Reference

[1] P. W. Anderson, Mater. Res. Bull. **8**, 153 (1973).
[2] R. Moessner and S. L. Sondhi, Prog. Theor. Phys. Suppl. **145**, 37 (2002).
[3] P. A. Lee, Science **321**, 1306 (2008).
[4] L. Balents, Nature (London) **464**, 199 (2010).
[5] M. P. Shores, E. A. Nytko, B. M. Bartlett, and D. G. Nocera J. Am. Chem. Soc. **127** 13462 (2005).
[6] P. Mendels, F. Bert, M. A. de Vries, A. Olariu, A. Harrison, F. Duc, J. C. Trombe, J. S. Lord, A. Amato, and C. Baines, Phys. Rev. Lett. **98,** 077204 (2007).
[7] Z. Hiroi, M. Hanawa, N. Kobayashi, M. Nohara, H. Takagi, Y. Kato, and M. Takigawa, J. Phys. Soc. Jpn. **70**, 3377 (2001).
[8] H. Yoshida, J. Yamaura, M. Isobe, Y. Okamoto, G. J. Nilsen, and Z. Hiroi, Nat. Commun. **3**, 860 (2012).
[9] Y. Okamoto, H. Yoshida, and Z. Hiroi, J. Phys. Soc. Jpn. **78** 033701 (2009).
[10] D. Boldrin, B. Fåk, E. Canévet, J. Ollivier, H. C. Walker, P. Manuel, D.D. Khalyavin, and A.S. Wills, Phys. Rev. Lett. **121**, 107203 (2018).
[11] H. Ishikawa, D. Nishio-Hamane, A. Miyake, M. Tokunaga, A. Matsuo, K. Kindo, and Z. Hiroi, Phys. Rev. Materials **3**, 064414 (2019).
[12] J. Schulenburg, J. Richter, and D. D. Betts, Acta Phys. Pol. A **97**, 971 (2000).
[13] D. Schmalfuß, P. Tomczak, J. Schulenburg, and J. Richter, Phys. Rev. B **65**, 224405 (2002).
[14] D. J. J. Farnell, R. Darradi, R. Schmidt, and J. Richter, Phys. Rev. B **84**, 104406 (2011).
[15] Y. Haraguchi, A. Matsuo, K. Kindo, and Z. Hiroi, Phys. Rev. B **98**, 064412 (2018).
[16] A. Almaz, M. Huvé, S. Colis, M. Colmont, A. Diana, and O. Mentré, Angew. Chem., Int. Ed. 51, 9393 (2012).
[17] D. Cave, F. C. Coomer, E. Molinos, H.-H. Klauss, and P. T. Wood, Angew. Chem. Int. Ed. **45**, 803 (2006)
[18] T. Fennell, J. O. Piatek, R. A. Stephenson, G. J. Nilsen, and H. M. Ronnow, J. Phys.: Condens. Matter **23**, 164201 (2011).
[19] S. J. Mills, A.R. Kampf, A.G. Christy, R.M. Housley, G.R. Rossman, R.E. Reynolds, J. Marty, Mineral. Mag. **78**, 1325 (2014).
[20] F. Izumi and K. Momma, Solid State Phenom. **130**, 15 (2007).
[21] K. Momma and F. Izumi, J. Appl. Crystallogr. **44**, 1272 (2011).
[22] T. Moriya and K. Yosida, Prog. Theor. Phys. **9**, 663 (1953).
[23] J. C. Bonner and M. E. Fisher, Phys. Rev. **135**, A640 (1964).
[24] G. Mennenga, L. J. de Jongh, W. J. Huiskamp, and J. Reedijk, J. Magn. Magn. Mater. **44**, 89 (1984).
[25] D. C. Dender, P. R. Hammar, D. H. Reich, C. Broholm, and G. Aeppli, Phys. Rev. Lett. **79**, 1750 (1997).
[26] D. C. Johnston, R. K. Kremer, M. Troyer, X. Wang, A. Klümper, S. L. Bud'ko, A. F. Panchula, and P. C. Canfield, Phys. Rev. B **61** 9558 (2000).
[27] P. Svoboda, P. Javorský, M. Diviš, V. Sechovský, F. Honda, G. Oomi, and A. A. Menovsky, Phys. Rev. B **63**, 212408 (2001).
[28] M. Shiroishi and M. Takahashi, Phys. Rev. Lett. **89**, 117201 (2002).
[29] J. B. Parkinson and J. C. Bonner, Phys. Rev. B **32**, 4703 (1985).
[30] C. P. Landee and M. M. Turnbull, J. Coord. Chem. **67**, 375

(2014).

[31] L. de Jongh and A. Miedema, Adv. Phys. **23**, 1 (1974).

[32] S.-H. Lee, C. Broholm, W. Ratcliff, G. Gasparovic, Q. Huang, T. H. Kim, and S.-W. Cheong, Nature (London) **418**, 856 (2002).

[33] S.-H. Lee, D. Louca, H. Ueda, S. Park, T. J. Sato, M. Isobe, Y. Ueda, S. Rosenkranz, P. Zschack, J. Iniguez, Y. Qiu, and R. Osborn, Phys. Rev. Lett. **93**, 156407 (2004).

[34] M. Q. Weng, D. N. Sheng, Z. Y. Weng, and R. J. Bursill, Phys. Rev. B **74**, 012407 (2006).

[35] R. Coldea, D. A. Tennant, A. M. Tsvelik, and Z. Tylczynski, Phys. Rev. Lett. **86**, 1335 (2001).

[36] R. Coldea, A. Tennant, and Z. Tylczynski, Phys. Rev. B **68**, 134424 (2003).

[37] D. Hirai, K. Nawa, M. Kawamura, T. Misawa, and Z. Hiroi, J. Phys. Soc. Jpn. **88**, 044708 (2019).

[38] D. Hirai, T. Yajima, K. Nawa, M. Kawamura, and Z. Hiroi, Inorg. Chem. **59**, 10025 (2020).

[39] R. Okuma, T. Yajima, T. Fujii, M. Takano, Z. Hiroi, J. Phys. Soc. Jpn. **87**, 093702 (2018).

[40] R. Okuma, M. Kofu, S. Asai, M. Avdeev, A. Koda, H. Okabe, M. Hiraishi, S. Takeshita, K. M. Kojima, R. Kadono, T. Masuda, K. Nakajima, and Z. Hiroi, Nat. Commun. **12**, 4382 (2021).

[41] R. Makuta and C. Hotta, arXiv:2110.06491.

[42] M. Moliner, I. Rousochatzakis, and F. Mila, Phys. Rev. B **83**, 140414(R) (2011).

[43] A.R. Kampf, S. J. Mills, R.M. Housley, J. Marty, Am. Mineral. **98**, 506 (2013).

[44] F. Olmi, A. Santucci, R. Trosti-Ferroni, Eur. J. Mineral **7** 1325 (1995)